\pdfoutput=1
\documentclass[acmsmall,screen]{acmart}
\setcopyright{none}
\copyrightyear{2024}
\acmYear{2024}
\acmDOI{}

\usepackage{fancyhdr}

\usepackage{tabularx,pgfplots,tikz,bussproofs,amsthm,soul}
\usepackage[frozencache]{minted}
\usepackage{xspace}
\usepackage{listings}

\pgfplotsset{compat=1.18}

\lstdefinelanguage{lama}{
keywords={box, unbox, string, sexp, true, false, fun, case, of, esac, let, in, eta, import, public, infix, infixl, infixr, at, before, after, syntax, local, var, if, then, else, fi, do, od, for},
sensitive=true,
commentstyle=\small\itshape\ttfamily,
keywordstyle=\textbf,
identifierstyle=\ttfamily,
basewidth={0.5em,0.5em},
columns=fixed,
mathescape=true,
fontadjust=true,
literate={->}{{$\to$}}3{=>}{{$\Rightarrow$}}3{=>>}{{$\Rightarrow$\hspace{-0.7em}$\Rightarrow$}}3,
morecomment=[s]{(*}{*)},
basicstyle=\small,
morestring=[b]"
}

\newcommand{\lama}{$\lambda\kern -.1667em\lower -.5ex\hbox{$a$}\kern -.1000em\lower .2ex\hbox{$\mathcal M$}\kern -.1000em\lower -.5ex\hbox{$a$}$\xspace}

\makeatletter
\@namedef{T1/zi4/m/it}{<->ssub*zi4/m/n}
\makeatother

\newcolumntype{L}{>{$\bf}l<{$}}

\title{A Relational Solver for Constraint-based Type Inference}

\author{Eridan Domoratskiy}
\email{eridan200@mail.ru}
\orcid{0009-0005-9020-3403}
\affiliation{%
  \institution{ITMO University}
  \city{St.~Petersburg}
  \country{Russia}
}

\author{Dmitry Boulytchev}
\email{dboulytchev@math.spbu.ru}
\orcid{0000-0001-8363-7143}
\affiliation{%
  \institution{St.~Petersburg State University}
  \city{St.~Petersburg}  
  \country{Russia}
}



\begin{document}

\settopmatter{printacmref=false}
\settopmatter{printfolios=true}
\renewcommand\footnotetextcopyrightpermission[1]{}
\pagestyle{fancy}
\fancyfoot{}
\fancyfoot[R]{miniKanren'24}
\fancypagestyle{firstfancy}{
  \fancyhead{}
  \fancyhead[R]{miniKanren'24}
  \fancyfoot{}
}
\makeatletter
\let\@authorsaddresses\@empty
\makeatother

\begin{abstract}
We present a \textsc{miniKanren}-based type inferencer for an educational
programming language with first-class functions, S-expressions, and
pattern-matching. The language itself is untyped which adds a certain
specificity to the problem and requires the employment of techniques
conventionally used in implicit/gradual typing settings. The presence of
polymorphic and recursive types poses a certain challenge when
implementing the inferencer in \textsc{miniKanren} and requires a number
of tricks, optimizations, and extensions to be used; we report on those
as well.
\end{abstract}





\maketitle
\thispagestyle{firstfancy}
\section{Introduction}

Type inference/checking/inhabitation is often considered as exemplary problems for relational programming.
The connection between them makes it possible to demonstrate its potential in expressing inverse computations
and utilizing the verifier-to-solver approach~\cite{SearchProblems}. However, the applicability of
this idea for realistic type systems is still a matter of discussion. While the approach works nicely
for toy type systems like STLC, for more complex type systems its direct application does not deliver
encouraging results. Even for a mild generalization of STLC, the Hindley-Milner type system~\cite{HM1,HM2},
its proper implementation in \textsc{miniKanren} relies on not-quite-relational tricks or even
a heavy machinery like \textsc{miniKanren}-in-\textsc{miniKanren} implementation~\cite{RelConv}.

In this paper we present preliminary results on using relational programming for implementing a
type inference for a realistic programming language with first-class functions, S-expressions, and
pattern-matching. This language, \lama~\cite{Lama}, has been used for a few years as an
educational language to teach compiler construction courses in a number of universities. While
not quite being the production-tier programming language, \lama still is rich enough to, first,
demonstrate the majority of relevant techniques in compiler construction domain, and, second, to
implement its own compiler.

An important feature of \lama is the lack of a type system. This brings in all well-known advantages and
drawbacks. The motivation for this work was to soften the latter by providing an automatic tool to
discover inconsistencies in \lama programs caused by incoherent usage of data, which is conventionally
done by means of a type system; thus, we call our approach ``type inference''. In a nutshell, in our
approach a set of constraints is extracted from a program, and the objective is to check if this
set of constraints is consistent. This approach is not entirely 
new~\cite{ConstrainedTypes}, and
the lack of an explicit type system just makes the setting similar to those involving
implicit types like polymorphic variants~\cite{PV} or gradual typing~\cite{Grad1, Grad2}. While
constraint extraction is done in a conventional syntax-directed way and implemented in a functional
language, the consistency check is performed relationally. However, as a naïve relational implementation
does not perform well, a number of refinements, optimizations and extensions to the vanilla \textsc{miniKanren}~\cite{ReasonedSchemer}
have been used.






The implementation is done in \textsc{OCaml} with \textsc{OCanren}~\cite{kosarev2018typed} utilized as a relational engine. 

\section{The \lama Programming Language}

The \lama programming language~\cite{Lama-impl} has been developed around 2018 by JetBrains Research as a supplementary
language for a compiler construction course. The compiler was originally written in \textsc{OCaml} but currently is
being bootstrapped. In this section we give an informal overview of the essential features of the language in order to
provide a context for a more formal description in the following sections.

In the context of this work it is important that \lama does not possess a conventional type system since the very objective
of its design is to provide a substrate for demonstrating the variety of runtime behaviors and relevant implementation
techniques. However, this flexibility comes at a high price since the compiler willingly accepts a lot of ill-formed
programs which then need to be debugged and fixed. Thus, the motivation for this work was to provide an optional
tool which would discover a certain class of inconsistencies in \lama programs. Unlike a conventional type checker
this tool would not reject programs but rather provide hints about potential problems, i.e. perform as
a static analyzer.

All values a \lama program operates with can be categorized as either integer numbers or references. The references
in turn can point to the structures of the following few shapes:

\begin{itemize}
\item strings (\lstinline[language=lama]|"this is a string"|);
\item S-expressions (\lstinline[language=lama]|Person ("John Doe", 1970)|);
\item tuples/arrays (\lstinline[language=lama]|[42, "is the", Answer]|);
\item closures (\lstinline[language=lama]|fun (x) {x}|).
\end{itemize}

Under the hood, all these shapes are implemented in a similar manner as arrays of values augmented with a small piece of
meta-information; ``fixnum'' representation is used to tell integers and references apart. This unicity of
representation makes it possible to manipulate the data generically. For example, a primitive ``\lstinline[language=lama]|length|''
can be used to request the number of immediate subvalues for a datum of any shape (which gives the number of
immediate components for a tuple, the length for a string, the number of arguments for an S-expression and
the number of captured variables for a closure), and ``\lstinline[language=lama]|$v$ [$i$]|'' allows
to read/write a certain immediate component of a datum $v$ by its integer index $i$.

The language is equipped with a simplistic pattern-matching, which allows for deconstructing the values in
a conventional manner; in addition the language of patterns makes it possible to discriminate on the \emph{shape} of a datum:

\begin{itemize}
\item ``\lstinline[language=lama]|#unbox|'' matches arbitrary integer;
\item ``\lstinline[language=lama]|#box|'' matches arbitrary reference;
\item ``\lstinline[language=lama]|#sexp|'' matches arbitrary S-expression;
\item ``\lstinline[language=lama]|#string|'' matches arbitrary string;
\item ``\lstinline[language=lama]|#fun|'' matches arbitrary closure.
\end{itemize}

All these features together make it possible to manipulate data in various ways, including inconsistent ones. Some
of these inconsistencies are relatively harmless. For example, given a function

\begin{lstlisting}[language=lama]
   fun size (l) {
     case l of
       Nil          -> 0
     | Cons (_, tl) -> 1 + size (tl)
     esac
   }
\end{lstlisting}

for a list length calculation the call ``\lstinline[language=lama]|size (1)|'' is invalid (and would not typecheck
in the majority of conventional typed languages). However, being still performed this call immediately results in
a runtime error, which makes it possible to identify and fix the bug.

There are, however, some inconsistencies which may require hours of code inspection and debugging. Consider,
for example, the following (artificial) function:

\begin{lstlisting}[language=lama]
   fun f ($\otimes$, x, y) {
     Array.lookup ($\otimes$,
       [ ["+", fun (x, y) {x + y}]
       , ["-", fun (x)    {x - y}]
       , ["*", fun (x, y) {x * y}] 
       ]
     ) (2 * x, 3 * y)
   }
\end{lstlisting}

This function gets a binary operator ``$\otimes$`` as a string and two integer values $x$ and $y$ and (presumably)
calculates the value $(2\times x)\otimes(3\times y)$ depending on what arithmetic operator ``$\otimes$'' actually
designates. However, there is a subtle and hard-to-discover typo here: the second argument of the function for
subtraction is missed, and in its body ``\lstinline|y|'' will be bound to the argument of the enclosing function.
Thus, instead of $2\times x - 3\times y$ the value of $2\times x - y$ will be calculated. Not only this
bug will hardly reveal itself by a runtime error, it may sometimes (but not always) lead to miscalculations
the reasons of which are hard to identify. At the same time a routine type check would discover the
error immediately.

\section{Type system}

In our work we treat the \emph{shapes} of data structures as their types. From a program we infer a set
of constraints for the shapes of values, and then check if the set of constraints is consistent. Since the shapes
are generic, the types are generic, too. For example, given a piece of code ``\lstinline[language=lama]|a [n.length]|``
we can not infer (as in strongly-typed languages) that ``\lstinline[language=lama]|a|'' is an array (or string);
we can only state that it is something from which a subvalue can be taken; similarly, the only constraint
we can infer (from this sample) for ``\lstinline[language=lama]|n|'' is that it is a reference. The lack of
used-defined types complicates the inference: for example, we have no way to infer recursive types other
than by reconstructing them from recursive functions; the same is true for polymorphism.

More precisely, we use three syntactic domains to define types:
\begin{center}
    \begin{equation*}
        \begin{array}{Llrl}
            Types & \texttt{T} & ::= & \mathcal{X} \mid \mathbb{Z} \mid \mathbb{S} \mid [\texttt{T}] \mid \mathcal{X}(\texttt{T}, ..., \texttt{T}) \sqcup ... \sqcup \mathcal{X}(\texttt{T}, ..., \texttt{T}) \\
            && \mid & \forall \mathcal{X}, ..., \mathcal{X}. ~ C \Rightarrow (\texttt{T}, ..., \texttt{T}) \rightarrow \texttt{T} \mid \mu \mathcal{X}. ~ \texttt{T} \\
            Constraints & C & ::= & \top \mid C \land C \mid Ind(\texttt{T}, \texttt{T}) \mid Call(\texttt{T}, \texttt{T}, ..., \texttt{T}, \texttt{T}) \\
            && \mid & Match(T, \Pi, ..., \Pi) \mid Sexp_\mathcal{X}(\texttt{T}, \texttt{T}, ..., \texttt{T}) \\
            Type patterns & \Pi & ::= & \_ \mid \texttt{T}@\Pi \mid [\Pi, ..., \Pi] \mid \mathcal{X}(\Pi, ..., \Pi) \\
            && \mid & \texttt{\#box} \mid \texttt{\#unbox} \mid \texttt{\#str} \mid \texttt{\#array} \mid \texttt{\#sexp} \mid \texttt{\#fun} \\
        \end{array}
    \end{equation*}
\end{center}

Non-parametric types $\mathbb{Z}$ and $\mathbb{S}$ stand for integers and strings respectively, for instance $42 : \mathbb{Z}$ or $\text{\lstinline[language=lama]|"text"|} : \mathbb{S}$.

Types of S-expressions have the form $\mathcal{X}_1(\texttt{T}_{1, 1}, ..., \texttt{T}_{1, n_1}) \sqcup ... \sqcup \mathcal{X}_n(\texttt{T}_{n, 1}, ..., \texttt{T}_{n, n_n})$, where $n$ is a
number of S-expression constructors, $\mathcal{X}_i$ is a tag of the constructor number $i$ and $\texttt{T}_{i, j}$ is a type of the $j$-th parameter for the $i$-th constructor.
For example, in the following code we may infer that $\text{\lstinline[language=lama]|x|} : A(\mathbb{Z}) \sqcup B(\mathbb{S})$:

\begin{lstlisting}[language=lama]
   var x = A (42);
   x := B ("text")
\end{lstlisting}

Note, in order to infer the types for S-expressions we need to track down all instantiations for all tags for the same type. Sometimes
the result can not be expressed in our type system since the same arguments for the same tag can require different types in different contexts.

The types of the form $[\texttt{T}]$ (for any type \texttt{T}) are homogeneous arrays with elements of type \texttt{T}, e.g. $\text{\lstinline[language=lama]|[1, 2, 3]|} : [\mathbb{Z}]$.
Heterogeneous arrays are forbidden, so arrays cannot be used as tuples, but S-expressions can: \lstinline[language=lama]|[1, "2", Three]| cannot be typed,
but $\text{\lstinline[language=lama]|Tuple (1, "2", Three)|} : Tuple\,(\mathbb{Z}, \mathbb{S}, Three)$.

Closures have types of the form $\forall \mathcal{X}_1, ..., \mathcal{X}_m. ~ C \Rightarrow (\texttt{T}_1, ..., \texttt{T}_n) \rightarrow \texttt{T}$, where $\mathcal{X}_i$
are bound type variables so we can type polymorphic functions (e.g. $\text{\lstinline[language=lama]|fun (x) \{ x \}|} : \forall a. ~ \top \Rightarrow (a) \rightarrow a$),
$C$ is a constraint over bound variables (like in \textsc{Haskell} but instead of type classes we use a closed set of predefined constraints), $\texttt{T}_i$ are types of
parameters and \texttt{T} is the type of result.

To be able to type recursive data structures like lists we additionally define a form of recursive types as $\mu \mathcal{X}. ~ \texttt{T}$, where $\mathcal{X}$ is a recursive
type variable. So we can infer $\text{\lstinline[language=lama]|xs|} : \mu a. ~ Nil \sqcup Cons\,(\mathbb{Z}, a)$ from the following code:

\begin{lstlisting}[language=lama]
   var xs = Nil ;
   xs := Cons (42, xs)
\end{lstlisting}

In this type system the type for the function ``\lstinline|size|'' given in the previous section can be specified as

\[
\forall a. ~ \top \Rightarrow (\mu b. ~ Nil \sqcup Cons(a, b)) \rightarrow \mathbb{Z}
\]

Here we say that the function accepts values of type $\mu b. ~ Nil \sqcup Cons(a, b)$ which is, as we mentioned before, a type for lists, and returns integer.

Type equality (denoted as $\texttt{T} \equiv \texttt{T}$ and used implicitly in inference rules) designates a syntactic equality w.r.t. recursive types unfolding.
In particular, this means that we do not work take into account the $\alpha$-equivalence of types yet (e.g. $\forall x_1. ~ \top \Rightarrow (x_1) \rightarrow x_1 \not\equiv \forall x_2. ~ \top \Rightarrow (x_2) \rightarrow x_2$). Recursive type unfolding stands for an operation that acts in the following manner: $\mu \mathcal{X}. ~ \texttt{T} \mapsto \texttt{T} [\mathcal{X} \mapsto \mu \mathcal{X}. ~ \texttt{T}]$, where type substitution is surrounded by brackets. This means that, e.g., $\mu x_1. ~ \mathbb{Z} \equiv \mu x_2. ~ \mathbb{Z}$ since $\mu x_1. ~ \mathbb{Z} \equiv \mathbb{Z} \equiv \mu x_2. ~ \mathbb{Z}$. This relation naturally expands to another syntactic domains.

Finally, we must mention that our system lacks a principal type. For example, in the context of the following program

\begin{lstlisting}
   var x;
   write (x[0])
\end{lstlisting}

the type of ``\lstinline|x|'' can be either an S-expression, or an array, or a closure, etc.

\section{Constraint system}

Now we describe our constraint system. First, there is a closed set of \emph{atomic} constraints: $\top$, $Ind$, $Sexp$, $Call$, and $Match$.

``$\top$'' is a vacuous constraint which is always satisfied.

As we mentioned before, we can not infer the exact type for a variable ``\lstinline[language=lama]|a|'' from a context ``\lstinline[language=lama]|a [i]|'' because it can be a string,
an array or an S-expression. To limit the set of possible types we use the constraint $Ind\,(\texttt{T}, \texttt{S})$ to express the fact that type \texttt{T} is a type of
containers with elements of type \texttt{S}. For instance, given a piece of code ``\lstinline[language=lama]|xs [i] := 42|'' and $\text{\lstinline[language=lama]|xs|} : a$ we infer the
constraint $Ind\,(a, \mathbb{Z})$.

As we mentioned before, we need to track all constructors of S-expression type instead of assigning some specific type eagerly. To achieve that, we use the constraints of the
form $Sexp_\mathcal{X}\,(\texttt{T}, \texttt{S}_1, ..., \texttt{S}_n)$ to express the fact that type \texttt{T} is a type of S-expression and one of its constructors
is $\mathcal{X}(\texttt{S}_1, ..., \texttt{S}_n)$. We write tag $\mathcal{X}$ as subscript, because it is known at the constraint generation time so we could
look on this form of constraints like on the family of forms indexed with different tags. As an example, given the following code, we say that the type
of ``\lstinline[language=lama]|x|'' (e.g. ``$a$'') must satisfy the constraints $Sexp_A\,(a, \mathbb{Z})$ and $Sexp_B\,(a, \mathbb{S})$ simultaneously:

\begin{lstlisting}[language=lama]
   var x = A (42) ;
   x := B ("text")
\end{lstlisting}

Similarly to S-expressions, when we identify that a certain type is a function we cannot say immediately what type variables are bound in this type and what
constraints should them satisfy. Thus, we use an atomic constraint of the form $Call\,(\texttt{T}, \texttt{S}_1, ..., \texttt{S}_n, \texttt{S})$ to express the fact that
values of type ``\texttt{T}'' must be callable with $n$ arguments of types $\texttt{S}_i$ and the type of result is \texttt{S}. For example, given an expression
``\lstinline[language=lama]|f (42, "text")|'' of type ``$b$'' and $\text{\lstinline[language=lama]|f|} : a$ we infer the constraint $Call\,(a, \mathbb{Z}, \mathbb{S}, b)$.

Finally, we use constraints of the form $Match\,(\texttt{T}, \Pi_1, ..., \Pi_n)$ and type patterns (denoted by $\Pi$) to express that the values of type \texttt{T} must be
matchable with patterns that correspond to given type patterns $\Pi_i$.

Since, as we saw previously, in some contexts we need types to satisfy multiple constraints at the same time, we introduce composite constraints in the form $C_1 \land C_2$.

To define what it means that ``constraints are satisfied'' we use a constraint entailment relation denoted as $C \Vdash C$. We say that ``$C_1$ implies $C_2$`` if relation
$C_1 \Vdash C_2$ holds. The ``$\Vdash$'' relation is defined via a conventional inference system shown in Fig.~\ref{fig:rules}; we denote by $\sigma\equiv[\mathcal{X}_i \mapsto \texttt{U}_i]$
the (simultaneous) substitution of types ``$\texttt{U}_i$'' for type variables ``$\mathcal{X}_i$'', and application of a substitution by juxtaposition.
Since we aren't talking about $Match$ constraints below, there aren't inference rules for them.

\begin{figure}
    \centering
    \begin{tabular}{cccc}
        \multicolumn{4}{c}{\fbox{$C \Vdash C$}} \\
        \\
        $C \Vdash C$
        &
        (C -- Refl)
        &
        $C \Vdash \top$
        &
        (C -- Top)
        \\ \\
        \multicolumn{3}{c}{
            \AxiomC{$C \Vdash C_1$}
            \AxiomC{$C \Vdash C_2$}
            \BinaryInfC{$C \Vdash C_1 \land C_2$}
            \DisplayProof
        }
        &
        (C -- And)
        \\ \\
        \AxiomC{$C_1 \Vdash C$}
        \UnaryInfC{$C_1 \land C_2 \Vdash C$}
        \DisplayProof
        &
        (C -- AndL)
        &
        \AxiomC{$C_2 \Vdash C$}
        \UnaryInfC{$C_1 \land C_2 \Vdash C$}
        \DisplayProof
        &
        (C -- AndR)
        \\ \\
        $C \Vdash Ind(\mathbb{S}, \mathbb{Z})$
        &
        (C -- IndString)
        &
        $C \Vdash Ind([\texttt{T}], \texttt{T})$
        &
        (C -- IndArray)
        \\ \\
        \multicolumn{3}{c}{
            $C \Vdash Ind(\mathcal{X}_1(\texttt{T}, ..., \texttt{T}) \sqcup ... \sqcup \mathcal{X}_n(\texttt{T}, ..., \texttt{T}), \texttt{T})$
        }
        &
        (C -- IndSexp)
        \\ \\
        \multicolumn{3}{c}{
            $\sigma \equiv [\mathcal{X}_1 \mapsto \texttt{U}_1, ..., \mathcal{X}_m \mapsto \texttt{U}_m]$
        }
        \\
        \multicolumn{3}{c}{
            \AxiomC{$C \Vdash C' \sigma$}
            \AxiomC{$\forall i \in [n]. ~ \texttt{T}_i \sigma \equiv \texttt{S}_i$}
            \AxiomC{$\texttt{T} \sigma \equiv S$}
            \TrinaryInfC{$C \Vdash Call(\forall \mathcal{X}_1, ..., \mathcal{X}_m. ~ C' \Rightarrow (\texttt{T}_1, ..., \texttt{T}_n) \rightarrow \texttt{T}, \texttt{S}_1, ..., \texttt{S}_n, \texttt{S})$}
            \DisplayProof
        }
        &
        (C -- Call)
        \\ \\
        \multicolumn{3}{c}{
            $\exists i \in [n]. ~ \mathcal{X}_i \equiv \mathcal{X} \land n_i \equiv m$
        }
        \\
        \multicolumn{3}{c}{
            \AxiomC{$\forall i \in [n]. ~ \mathcal{X}_i \equiv \mathcal{X} \land n_i \equiv m \implies \forall j \in [m]. ~ \texttt{T}_{i, j} \equiv \texttt{S}_j$}
            \UnaryInfC{$C \Vdash Sexp_\mathcal{X}(\mathcal{X}_1(\texttt{T}_{1, 1}, ..., \texttt{T}_{1, n_1}) \sqcup ... \sqcup \mathcal{X}_n(\texttt{T}_{n, 1}, ..., \texttt{T}_{n, n_n}), \texttt{S}_1, ..., \texttt{S}_m)$}
            \DisplayProof
        }
        &
        (C -- Sexp)
        \\
    \end{tabular}
    \caption{Constraint entailment inference rules}
    \label{fig:rules}
\end{figure}

\section{Constraint solver}

In this section we describe the peculiarities of the relational solver implementation for our constraint system. Additionally
we describe some \textsc{OCanren} modifications which we used to implement the solver.

We start from introducing the simple \textsc{miniKanren} primitives that allow
to distinguish between free and bound logic variables,
then we describe their useful applications in our solver.
Next, we present a possible way to deal with recursive terms
without recursive substitutions support in  relational engine.
Finally, we share our approach to relational query construction for given domain-specific task.
On this way, we additionally highlight some useful well-known relational
programming tricks that help us to improve the solver.

\subsection{Term shape check helpers}

Each type of S-expression is fully characterized by the set of its constructors and, for each constructor, the number and types of its arguments. Similarly,
a function type is fully characterized by the number and types of a function arguments. Thus,
when we check (or solve) constraints of the form $Sexp$ we need to require that the list of constructors includes given label and an
associated list of types corresponds to the given one, and similarly for the constraints of the form $Call$. 
In practice this means that we need to iterate over the list of constructors which
in relational programming implies the synthesis of list if it isn't ground. In particular, when we check that some element is included in a list
we generate all possible lists which include given element. The usage of wildcard variables~\cite{kosarev2022wildcard} can prune the search
space in our case but this is still not enough.

To address this issue we define new primitives: \texttt{is\_var} and \texttt{is\_not\_var} which check if given term is a variable in
the current state or not. Here is the implementation for \textsc{OCanren}:

\begin{listing}[H]
    \begin{minted}{OCaml}
   let check_is_var ({env; subst} : State.t) (x : 'a ilogic) : bool =
     if Env.is_var env x
     then Env.is_var env @@ Subst.shallow_apply env subst x
     else false

   let is_var (x : 'a ilogic) : goal = fun st ->
     if check_is_var st x then Stream.single st else Stream.nil

   let is_not_var (x : 'a ilogic) : goal = fun st ->
     if check_is_var st x then Stream.nil else Stream.single st
    \end{minted}
    \caption{The implementation of \texttt{is\_var} and \texttt{is\_not\_var} primitives for \textsc{OCanren}}
\end{listing}

In the function \texttt{check\_is\_var} we use a new method of module \texttt{Subst} named \texttt{shallow\_apply}. Since \textsc{OCanren}
internally uses a triangular form of substitutions the regular method \texttt{Subst.apply} applies substitutions in multiple steps, but we don't
need all of them to distinguish variables and non-variable terms. More precisely, we need just one call of the internal method \texttt{walk}
that applies a substitution to the given variable until getting non-variable term:

\begin{listing}[H]
    \begin{minted}{OCaml}
   let shallow_apply env subst x =
     match Term.var x with
     | Some v -> begin
       match walk env subst v with
       | WC v | Var v -> Obj.magic v
       | Value x      -> Obj.magic x
       end
     | None -> x
    \end{minted}
    \caption{Implementation of \texttt{Subst.shallow\_apply}}
\end{listing}

\subsection{Pruning the search space for function types}

Using the primitives \texttt{is\_var} and \texttt{is\_not\_var} we can implement a (non-relational) optimization for solving the constraints
of the form $Call(\texttt{T}, \texttt{S}_1, ..., \texttt{S}_n, \texttt{S})$.
Given a non-variable function type as ``\texttt{T}'' we allow to apply the (C~--~Call) rule straightforwardly.
But if ``\texttt{T}'' is a free logic variable, relational solver assumes that ``\texttt{T}'' is a term of the form
``$\forall \mathcal{X}_1, ..., \mathcal{X}_m. ~ C \Rightarrow (\texttt{T}_1, ..., \texttt{T}_n) \rightarrow \texttt{T}$'';
or in terms of implementation, ``\mintinline{OCaml}|TArrow (fxs, fc, fts, ft)|''
where ``\texttt{fxs}'' is a list of \textit{bound variables} ($\mathcal{X}_1, ..., \mathcal{X}_m$) and ``\texttt{fc}'' is a \textit{bound constraint} ($\mathcal{C}$).
Assuming that all possible function types are given to us in constraints,
we can shrink the search space by forbidding the
generation of complex function types. To achieve this we simply check
if ``\texttt{fxs}'' and ``\texttt{fc}'' are free logic variables and in this case require them to be empty:

{\parindent0pt\tt\obeyspaces\obeylines\def {\mbox{\space}}
   \textcolor{gray}{(* ... *)}
   \& \{ is\_var fxs \& fxs == [] | is\_not\_var fxs \}
   \& \{ is\_var fc  \& fc  == [] | is\_not\_var fc  \}
   \textcolor{gray}{(* ... *)} %
}

This piece of code is given in the syntax of \textsc{OCanren} syntactic extension for \textsc{OCaml}. Note, in the solver we represent composite
constraints as lists of atomic ones.

To show how this optimization really affects the solving process suppose that we need to solve a constraint of the form $Call\,(\texttt{T}, \mathbb{Z}, \mathbb{S})$.
If \texttt{T} is some non-variable term we act as usual. But if \texttt{T} is a free variable we generate only \emph{one} branch where $\texttt{T} \equiv \forall \varnothing. ~ \top \Rightarrow (\mathbb{Z}) \rightarrow \mathbb{S}$ (or just $(\mathbb{Z}) \rightarrow \mathbb{S}$). Without this optimization we would generate a variety of function types with all possible
bound variable lists and constraints; but in practice, when we reach $Call$ with free function type, there are only two possibilities: it could be any function type that
satisfies given constraints (including that we forcefully set to empty) or it might be some specific type that we probably will be unable to find in an adequate time.

As a reader may notice, given approach could prevent us from finding a correct solution in some cases when we don't know the type of function at the moment when $Call$ is being solved
but will encounter it in the future, when the values we emptied could become established. We address this issue below.

\subsection{Pruning the search space for S-expression types}

As it was mentioned above, we can experience problems with over-generation of S-expression types while solving constraints. Using the new primitives we can require to generate
the elements of lists in the same order as the constraints are being solved (i.e. forbid permutations of lists), but we still have too much branches in the search tree as long as we
don't limit the length of generated lists.

In order to address this problem we initially count all possible constructors of S-expression types and maximal number of their arguments. The number of
constructors allows us to limit the length of lists of constructors while the number of arguments allows us to limit the length of lists of arguments.
The latter is not needed for solving $Sexp$ constraints as long as we have the exact lists of types, but helps us in dealing with the rule (C~--~IndSexp) where we don't have them.
The implementation of this approach is shown in Listing~\ref{lst:sexp}.

Additionally, during the counting, we replace all symbolic labels with unique numeric identifiers w.r.t. the number of arguments of constructors. This allows us to simplify the code of the
solver and optimize it so we don't need to check the number of elements of lists because it becomes explicitly specified. For example, given a constraint $Sexp_{Cons}(a, \mathbb{Z})$
we replace it with $Sexp_x(a, \mathbb{Z})$, where ``$x$'' stands for a unique encoding of constructor \lstinline[language=lama]|Cons| with exactly two arguments.

The operator ``\texttt{=\textasciitilde\textasciitilde=}'' used in Listing~\ref{lst:sexp} stands for relational implementation of type equality w.r.t. recursive type unfolding
specifically for lists of types. The necessity of this special relation instead of the default ``\texttt{==}'' is discussed below.

As a result, given a constraint $Sexp_{Cons}\,(a, \mathbb{Z})$ we won't generate all lists that contain the constructor $Cons\,(\mathbb{Z})$, but only
$\langle Cons(\mathbb{Z}) \rangle, \langle Cons(\mathbb{Z}), c_2 \rangle, ..., \langle Cons(\mathbb{Z}), c_2, ..., c_n \rangle$, where ``$n$'' is a maximal
number of constructors and ``$c_i$'' are logic variables for possible other constructors. Given a constraint $Sexp_{Cons}(Nil \sqcup c_1 \sqcup ... \sqcup c_n, \mathbb{Z})$
we will generate only one branch with $c_1 \equiv Cons(\mathbb{Z})$, etc.

\begin{listing}
    \begin{flushleft}
        {\tt\obeyspaces\obeylines\def {\mbox{\space}}
\textcolor{purple}{let} sexp\_x\_hlp x xs ts : goal =
    \textcolor{purple}{let} max\_length = !sexp\_max\_length \textcolor{purple}{in}
    \textcolor{purple}{let} check\_n n = \textcolor{purple}{if} n > max\_length \textcolor{purple}{then} failure \textcolor{purple}{else} success \textcolor{purple}{in}
~
    \textcolor{gray}{(* require that xs doesn't contain label x *)}
    \textcolor{purple}{let rec} not\_in\_tail n xs = \textcolor{purple}{let} n' = n + \textcolor{teal}{1} \textcolor{purple}{in} \textcolor{purple}{ocanren} \{ check\_n n \&
        \{ xs == []
        | \textcolor{purple}{fresh} x', xs' \textcolor{purple}{in} xs == (x', \_) :: xs' \& x =/= x' \& not\_in\_tail n' xs'
        \}
    \} \textcolor{purple}{in}
~
    \textcolor{gray}{(* require that xs contains exactly one label x with correct types *)}
    \textcolor{purple}{let rec} hlp n xs = \textcolor{purple}{let} n' = n + \textcolor{teal}{1} \textcolor{purple}{in} \textcolor{purple}{ocanren} \{ check\_n n \&
        \textcolor{purple}{fresh} x', ts', xs' \textcolor{purple}{in} xs == (x', ts') :: xs' \&
            \{ x == x' \& ts =\textasciitilde\textasciitilde= ts' \& not\_in\_tail n' xs'
            | is\_not\_var x' \& x =/= x' \& hlp n' xs'
            \}
    \} \textcolor{purple}{in}
~
    hlp \textcolor{teal}{0} xs %
        }
    \end{flushleft}
    \caption{The solver for the (C~--~Sexp) constraint entailment rule}
    \label{lst:sexp}
\end{listing}

\subsection{Partial support for recursive terms in \textsc{OCanren}}

At this moment \textsc{OCanren} doesn't support unification of recursive terms and uses occurs check as a guard to prevent generation of recursive substitutions.
In order not to break the soundness and completeness of the search we decided not to change the main algorithm of unification now and instead implemented
another approach to provide a partial support of recursive terms.

Our approach is based on the idea of occurs check utilization. We present ``occurs hooks''~--- a mechanism that allows users to associate a programmatic hook
with logic variable which is invoked when occurs check fails. This hook gives a way to suggest an alternative solution for unification instead of failing.
When occurs hook suggests an alternative term, we call occurs check for this term again, now with occurs hooks disabled to prevent infinite looping.

Since we work with a typed embedding of \textsc{miniKanren} where types are erased at runtime, we cannot use generic hooks which apply for
every logic variable because we cannot distinguish the types of ``occurred'' variables in runtime. Because of this, we allows users to register typed
hooks on particular variables when we able to use \textsc{OCaml} type system to ensure the type soundness.

In an ideal world we would like to associate occurs hooks with arbitrary terms. This would allow us to ``replace'' already partially unified terms with
suggested ones. But in the reality it isn't clear what to do when occurs check raises an error. First, it is not trivial to determine occurs hook that
must be called because the ``occurred'' variable could appear in different terms with different hooks. Further, even if we've got some suggestion for the term,
how to perform this replacement? It could be some kind of unnatural rollback in time followed by a bunch of different problems we are currently not ready
to deal with. So in our implementation occurs hooks may be registered only for variables.

The next question is what to do with occurs hooks when a variable is unified with a non-trivial term. We would like to ``reassign'' occurs hook to the variables
of this term but at runtime we don't know the types of variables. We could provide a user an ability to ``teach'' the solver how to reassign the hooks but it will
cause the slowdown of the solver since unification is a really frequent event in the search. But if we will not do anything about the hooks of unified variables it
may cause unnecessary memory consumption, so we decided to clear the occurs hooks storage after every unification with no respect to variables they associated with.
It means that we need to setup occurs hooks on the interesting variables immediately before the unification but this isn't a problem since we really need
to do this because of the previously discussed reasons.

Now, when we discussed the interface implementation details, we are ready to present the implementation for \textsc{OCanren}:

\begin{minted}{OCaml}
exception Occurs_check
type term_vars = { get: 'a. int -> 'a ilogic }

val bind_occurs_hook : 'a -> ('a, 'b) Reifier.t
                    -> (term_vars -> int -> 'b -> 'a) -> goal
\end{minted}

As we can see, the new primitive \texttt{bind\_occurs\_hook} accepts a term of type ``\texttt{'a}'', a reifier~\cite{kosarev2018typed} for a type ``\texttt{'b}''
and an occurs hook and returns a goal. Occurs hook's type looks quite strange: it is a function of a ``bag of variables'' of type \texttt{term\_vars}, the id of occurred
variable (of type \texttt{int}), and a reified term which caused the occurs check to fail. The most weird thing here is the first parameter that we call a ``bag of variables''.
As shown in the listing, it is just a polymorphic function from a logic variable id to an injected term of some arbitrary type.

The problem here is a possible type unsoundness that \texttt{term\_vars} causes. We need this because the result of reification erases the real logic variables and provides
only their ids and some additional information but for the construction of suggested term we need to ``revert'' reification with the old logic variables. Another approach to
achieve this would be to preserve the source logic variables in reified terms but this would involve a lot of work that may be done in future in order to fix the
current implementation.

The internal implementation of occurs hooks is straightforward: given a state we collect registered occurs hooks and pass them to the unification procedure. When occurs check fails
we just lookup registered hooks and call them. The \texttt{bind\_occurs\_hook} goal checks that given term is a variable and produces a state extended with given occurs hook.

\subsection{Recursive types introduction and elimination}

Our approach to work with recursive types is to prevent them from generation by any means except for occurs hooks. To achieve this we write the relational part
of the solver like there are no recursive types, but use a helper that allows to perform an unfolding in cases when recursive type is already introduced:

{\parindent0pt\tt\obeyspaces\obeylines\def {\mbox{\space}}
\textcolor{purple}{let} unmu t t' = \textcolor{purple}{ocanren}
    \{ is\_var t \& t == t'
    | is\_not\_var t \&
        \{ t =/= \textcolor{blue}{TMu} (\_, \_) \& t == t'
        | \textcolor{purple}{fresh} x, s \textcolor{purple}{in} t == \textcolor{blue}{TMu} (x, s) \& subst\_t [(x, t)] s t'
        \}
    \} %
}

Here we check if given type \texttt{t} is a logic variable and in this case assume that it isn't a recursive type. Otherwise, we check is it a
recursive type (here ``\textcolor{blue}{\tt TMu}'' is a constructor of recursive type term
and ``\texttt{\_}'' stands for a wildcard variable) and apply an unfolding substitution
if it is (``\texttt{subst\_t}'' is a relational goal that applies the given
substitution to the given term but we don't discuss it's implementation details here).
Note, the current implementation of \textsc{OCanren} syntactic extension
does not support wildcard variables so we added it separately.

As we discussed above, occurs hooks must be registered immediately before the unification, so we need an extra operator ``\texttt{=\textasciitilde=}''
besides conventional unification. Its implementation is straightforward except for some optimizations and is shown in Listing~\ref{lst:eq}.
This relation naturally scales to constraints and lists of types (``\texttt{=\textasciitilde\textasciitilde=}'').

Implementation of the aforementioned function \texttt{set\_occurs\_hook\_t} is straightforward but requires some additional code that reverses
reification (which is hidden behind the \texttt{logic\_t\_to\_injected} function), so we will show only a high-level part of the implementation in Listing~\ref{lst:occurs-hook}.
Here we just replace logic variable \texttt{v} with the type variable by extending given bag of variables and wrap the whole term in a recursive type constructor.

\begin{listing}
    \begin{flushleft}
        {\tt\obeyspaces\obeylines\def {\mbox{\space}}
\textcolor{purple}{let rec} eq\_t t t' = \textcolor{purple}{ocanren}
    \{ t == t'
    | t =/= t' \&
        \{ is\_var     t \& is\_var     t' \& t == t'
        | is\_var     t \& is\_not\_var t' \& set\_occurs\_hook\_t t  \& t == t'
        | is\_not\_var t \& is\_var     t' \& set\_occurs\_hook\_t t' \& t == t'
        | is\_not\_var t \& is\_not\_var t' \&
            \{ t == \textcolor{blue}{TName} \_ \& t == t'
            | t == \textcolor{blue}{TInt}    \& t == t'
            \textcolor{gray}{(* ... *)}
            | \{ \textcolor{purple}{fresh} x, t1, t1' \textcolor{purple}{in} t == \textcolor{blue}{TMu} (x, t1)
                \& t' == \textcolor{blue}{TMu} (x, t1') \& eq\_t t1 t1' \}
            | \{ \textcolor{purple}{fresh} t1 \textcolor{purple}{in} t == \textcolor{blue}{TMu} (\_, \_)
                \& t' =/= \textcolor{blue}{TMu} (\_, \_) \& unmu t t1 \& eq\_t t1 t' \}
            | \{ \textcolor{purple}{fresh} t1' \textcolor{purple}{in} t =/= \textcolor{blue}{TMu} (\_, \_)
                \& t' == \textcolor{blue}{TMu} (\_, \_) \& unmu t' t1' \& eq\_t t t1' \}
            \}
        \}
    \} %
        }
    \end{flushleft}
    \caption{Equality of types relation implementation}
    \label{lst:eq}
\end{listing}

\begin{listing}
    \begin{flushleft}
        {\tt\obeyspaces\obeylines\def {\mbox{\space}}
\textcolor{purple}{let} occurs\_hook\_t vars v =
    \textcolor{purple}{let} get\_var u = \textcolor{purple}{if} v = u \textcolor{purple}{then} \textcolor{blue}{Obj}.magic @@ tName v \textcolor{purple}{else} vars.get u \textcolor{purple}{in}
    \textcolor{purple}{fun} t -> tMu !!v @@ logic\_t\_to\_injected \{ get = get\_var \} t
~
\textcolor{purple}{let} set\_occurs\_hook\_t t = bind\_occurs\_hook t reify\_t occurs\_hook\_t %
        }
    \end{flushleft}
    \caption{Implementation of the function \texttt{set\_occurs\_hook\_t}}
    \label{lst:occurs-hook}
\end{listing}

\subsection{Constraint solving order and runtime scheduling}

A straightforward approach of solving constraints left-to-right works poorly. To address this issue we use a runtime scheduling during the search.

The first step is to rewrite the implementation of ``$\Vdash$'' relation from a na\"{i}ve recursive to a tail-recursive form. This allows us to deal
with all of due to be solved constraints, even if some of them are being added as a result of solving another constraints (as it happens in the (C~--~Call) rule).

Now we became capable to picking any of planned constraints to solve them out-of-order. To control the order of the search we calculate the
weights of all the constraints and pick a constraint with the minimal weight. The weight depends on the form of
constraint arguments (e.g. $Call$ with variable function picked only when there aren't any other constraints in queue).

\subsection{Invoking the solver}

To run the solver we need to formulate relational query. While in \textsc{OCanren} interface there is only one function \texttt{run} that could make queries
with statically known number of parameters we need to prepare our constraints to solve them for the all free type variables. To achieve this we formulate a
query for one variable that represents a list of all interesting variables. The list is being built while preparing the constraints and injecting them
into an \textsc{OCanren} internal representation using the builtin primitive named \texttt{call\_fresh}:

\begin{minted}{OCaml}
val call_fresh : ('a ilogic -> goal) -> goal
\end{minted}

As we can see, this function doesn't allow us to just get a fresh variable, but provides an interface continuation-passing style.
Since we need to recursively traverse given constraints it isn't convenient to use it in a straightforward manner, so we use CPS monad
with the \textsc{OCaml} binding operator syntax. The implementation and an example of usage is shown in Listing~\ref{lst:cps}.

\begin{listing}
    \begin{minted}{OCaml}
   module Monad = struct
     type 'a t = ('a -> goal) -> goal

     let return (x : 'a) : 'a t = fun f -> f x
     let ( >>= ) (m : 'a t) (k : 'a -> 'b t) : 'b t =
        fun f -> m (fun a -> k a f)

        module Syntax = struct
            let ( let* ) m k = m >>= k
        end
   end

   (* ... *)

   let rec inject_list f = function
   | [] -> M.return @@ List.nil ()
   | x :: xs ->
       let* x = f x in
       let* xs = inject_list f xs in
       M.return @@ List.cons x xs
    \end{minted}
    \caption{Implementation of CPS monad in \textsc{OCaml}}
    \label{lst:cps}
\end{listing}

As an initial continuation we pass a function that configures our solver (e.g. sets the maximum number of constructors in S-expression types) and performs relational query.

\section{Evaluation}

To evaluate our typechecker we used the set of \lama compiler tests. The majority of tests were successfully typechecked
but some drawbacks were discovered, too.

\begin{figure}
    \centering
    \begin{tikzpicture}[scale=2]
        \node at (-2, 0) {$\sigma_0$};
        \node (xyz) at (0, 0) {$\varnothing$};

        \node at (-2, -0.5) {$Sexp_A(x, \mathbb{Z})$};
        
        \node at (-2, -1) {$\sigma_1$};
        \node (x1yz) at (-1, -1) {$x \mapsto A(\mathbb{Z})$};
        \node (x2yz) at (1, -1) {$x \mapsto A(\mathbb{Z}) \sqcup x'$};
        
        \node at (-2, -1.5) {$Sexp_B(y, \mathbb{S})$};
        
        \node at (-2, -2) {$\sigma_2$};
        \node (xy1z) at (-1, -2) {$y \mapsto B(\mathbb{S})$};
        \node (xy2z) at (1, -2) {$y \mapsto B(\mathbb{S}) \sqcup y'$};

        \node at (-2, -2.5) {$Sexp_A(z, \mathbb{Z})$};
        
        \node at (-2, -3) {$\sigma_3$};
        \node (xyz1) at (-1, -3) {$z \mapsto A(\mathbb{Z})$};
        \node (xyz2) at (1, -3) {$z \mapsto A(\mathbb{Z}) \sqcup z'$};

        \draw[->] (xyz) -- (x1yz);
        \draw[->] (xyz) -- (x2yz);
        
        \draw[->] (x1yz) -- (xy1z);
        \draw[->] (x1yz) -- (xy2z);
        \draw[->] (x2yz) -- (xy1z);
        \draw[->] (x2yz) -- (xy2z);
        
        \draw[->] (xy1z) -- (xyz1);
        \draw[->] (xy1z) -- (xyz2);
        \draw[->] (xy2z) -- (xyz1);
        \draw[->] (xy2z) -- (xyz2);
    \end{tikzpicture}
    \caption{Solutions tree for several different $Sexp$ constraints}
    \label{fig:exp}
\end{figure}
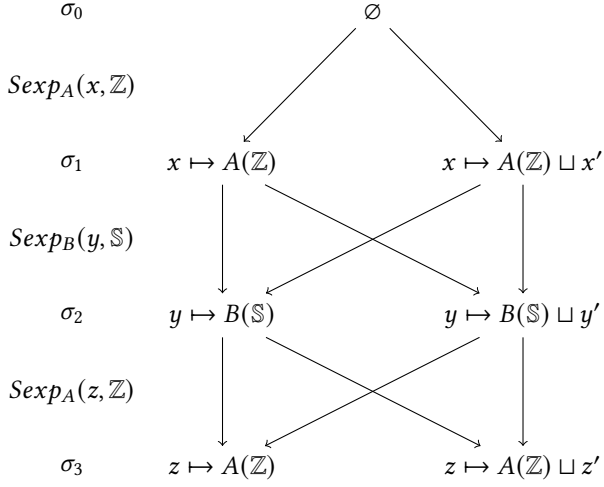

As we mentioned before when we solve the constraints of the form $Sexp$ on a logic variable we generate $n$ lists of lengths $1, 2, ..., n$,
where $n$ is the maximal number of S-expression constructors in the program. As a result when we have several $Sexp$ constraints on 
different logic variables we generate $O\,(n^m)$ branches in the search tree, where $m$ is the number of constraints. To demonstrate this,
assume we have the following constraints with two possible constructors:

\[Sexp_A(x, \mathbb{Z}) \land Sexp_B(y, \mathbb{S}) \land Sexp_A(z, \mathbb{Z}),\]

where $x, y, z$ are logic variables. The search tree will look like as shown in Fig.~\ref{fig:exp}. On this graph we use  additional
substitutions as nodes and a number of evaluation steps as edges. Every path expresses some path in the search tree,
so we got for our small example about $2^3$ branches, but we really need only the first of them. This problem could be solved by another
representation of S-expression types, but it requires further research.

Another problem which was discovered concerns occurs hooks: they generate not fully folded recursive types. For example, we could get a
type $Nil \sqcup Cons(\mathbb{Z}, \mu a. ~ Nil \sqcup Cons(\mathbb{Z}, a))$ instead of a simpler one $\mu a. ~ Nil \sqcup Cons(\mathbb{Z}, a)$.
This drawback cannot be solved without better way of handling recursive types.

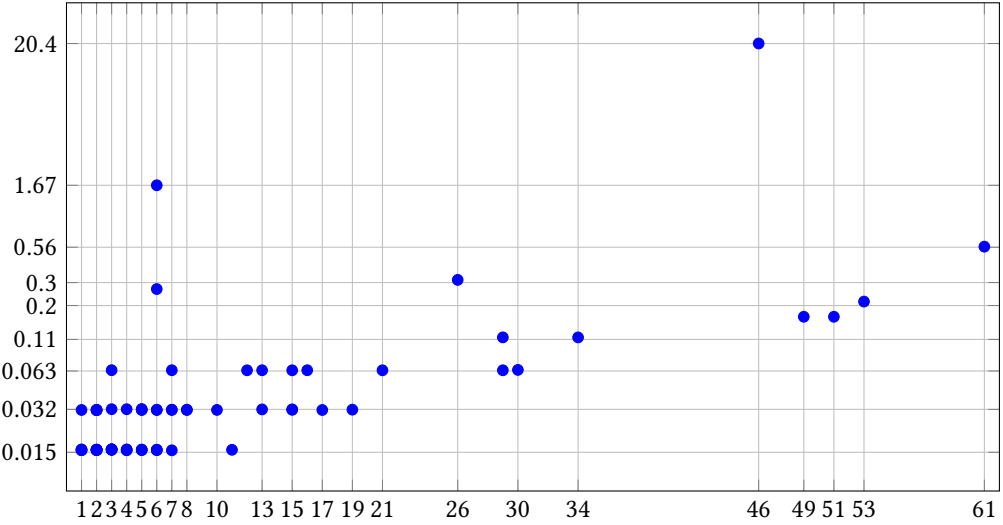
\begin{figure}
    \centering
    \begin{tikzpicture}
        \begin{axis}[
            width=14cm,
            height=8cm,
            xtick={1,2,3,4,5,6,7,8,10,13,15,17,19,21,26,30,34,46,49,51,53,61},
            xmin=0,
            xmax=62,
            xmajorgrids,
            ymode=log,
            ytick={0.015,0.032,0.063,0.11,0.2,0.3,0.56,1.67,20.4},
            ymajorgrids,
            log ticks with fixed point
        ]
            \addplot[only marks,color=blue] table {time_all.csv};
        \end{axis}
    \end{tikzpicture}
    \caption{Elapsed time for typecheck of tests}
    \label{fig:time-all}
\end{figure}

The elapsed time for the correctly typed tests is shown in Fig.~\ref{fig:time-all}. The unit for the axis X is the total number of solved constraints,
for the axis Y~--- the elapsed time for the test. Blue circles are tests, so we can see that the elapsed time doesn't linearly depend on the number of
the constraints.

As a number of solved constraints we use the number of solved constraints in the finished branch.
This, of course, does not take into account the contribution of the complexity of the constraints.

\begin{figure}
    \centering
    \begin{tikzpicture}
        \begin{loglogaxis}[
            width=14cm,
            height=8cm,
            xtick=data,
            ytick={0.015,0.032,0.063,0.11,0.2,0.3,0.56,1.67,20.4},
            xmajorgrids,
            ymajorgrids,
            log ticks with fixed point
        ]
            \addplot[only marks,mark=triangle*,color=red] table {time_sexp.csv};
        \end{loglogaxis}
    \end{tikzpicture}
    \caption{Elapsed time for typecheck of tests by number of $Sexp$ branches}
    \label{fig:time-sexp}
\end{figure}
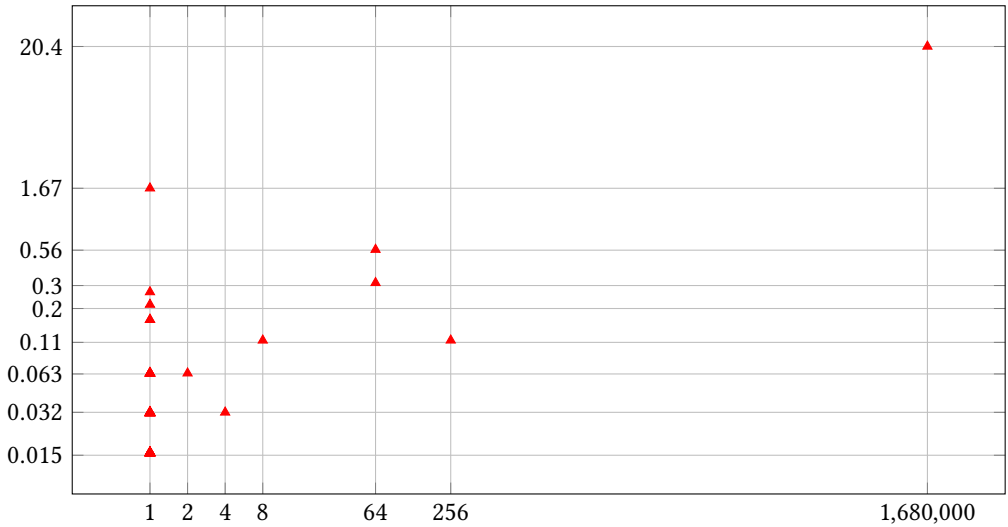

As we can see, we have an abnormal amount of time for the test with 46 constraints. We try to explain this anomaly by the next
plot in the Fig.~\ref{fig:time-sexp}. Here on the X axis is the $n^m$, where $n$ is the maximal number of constructors in S-expression
type and $m$ is the number of different logic variables that appear in the first place in the $Sexp$ constraint.

In this specific test, that runs for about 20 seconds, we have $6^8 = 1679616$ branches which causes so high time consumption.
As a positive result, we demonstrate that the time of evaluation highly depends only on a number of $Sexp$ constraints and this
could be optimized.

\subsection{Examples}

Here we present and comment on some concrete examples which type check.

The first example is pretty simple for our typechecker:

\begin{lstlisting}[language=lama]
   var n, x, i;

   fun sort (x) {
     var i, j, y, n = x.length;
  
     for i := 0, i < n, i := i + 1 do
       for j := i + 1, j < n, j := j + 1 do
         if x[j] < x[i] then
           y    := x[i];
           x[i] := x[j];
           x[j] := y
         fi
       od
     od;
     x
   }

   n := read ();
   x := [10, 9, 8, 7, 6, 5];
   x := sort (x);

   for i := 0, i < x.length, i := i + 1 do
     write (x[i])
   od
\end{lstlisting}

All constraints are $Call$'s and $Ind$'s, so we have no problems with solving them.

The next example is more interesting because of S-expression usage:

\begin{lstlisting}[language=lama]
   var x, y, i;

   fun f (x) {
     case x of
       Nil                               -> write (0)
     | Cons (_, Nil)                     -> write (1)
     | Cons (_, Cons (_, Nil))           -> write (2)
     | Cons (_, Cons (_, Cons (_, Nil))) -> write (3)
     | _                                 -> write (4)
     esac
   }
 
   x := read ();
   y := Nil;

   for i := 0, i < 10, i := i + 1 do
     f (y);
     y := Cons (i, y)
   od
\end{lstlisting}

Here we may notice that ``\lstinline[language=lama]|y|'' is initialized by assigning the ``\lstinline[language=lama]|Cons (i, y)|'' so we really
do have recursive type here. The constructed value is passed to the function ``\lstinline[language=lama]|f|'' that uses pattern matching to work
with this value of recursive type.

As we mentioned before we have some problems with the inference of the ``smallest'' recursive types, so in this example the type of ``\lstinline[language=lama]|y|''
is inferred as $Nil \sqcup Cons(\mathbb{Z}, \mu a. ~ Nil \sqcup Cons(\mathbb{Z}, a))$ instead of equivalent smaller one~--- $\mu a. ~ Nil \sqcup Cons(\mathbb{Z}, a)$.

When we say that we capable of working with recursive types we don't exclude recursive function types. The next example is typed correctly and it is a
really correct program that uses a recursive type of function:

\begin{lstlisting}[language=lama]
   var f = fun () {
     fun f (x) {
       fun () {
         write (x) ;
         f (x + 1)
       }
     }

     f (0)
   } () ;

   f () () () ()
\end{lstlisting}

The type inferred for ``\lstinline|f|'' is

\[
\forall a. ~ Call(\mu b. ~ \forall \varnothing. ~ \top \Rightarrow (\mathbb{Z}) \rightarrow \forall c. ~ Call(b, \mathbb{Z}, c) \Rightarrow () \rightarrow c, \mathbb{Z}, a) \Rightarrow () \rightarrow a
\]

and it can be actually further simplified.

In contrast to the last example, we can show the code, that doesn't type:

\begin{lstlisting}[language=lama]
   var x = [fun () { x [0] () }] ;
  
   x [0] ()
\end{lstlisting}

This is not a drawback of the described approach, but a little interesting example of how recursive types could make solvers loop without any special handling.
In this code the last line generates a constraint like

\[Call(\mu a. ~ \forall b, c. ~ Ind([a], b) \land Call(b, c) \Rightarrow () \rightarrow c, t_1).\]

When we use the rule (C~--~Call) to solve it we produce new constraints

\[Ind([\mu a. ~ \forall b, c. ~ Ind([a], b) \land Call(b, c) \Rightarrow () \rightarrow c], t_2) \land Call(t_2, t_1).\]

And on the next step, when we solve the constraint of form $Ind$, we returns back to the initial state. In other words, we don't move forward
but stay on the same place unable to do something else. To address this problem we need to solve this constraint in a more general manner.

\section{Conclusion and future work}

We presented a number of optimizations which could be helpful in making solvers for constraint-based type inference
applicable: pruning search space for some data structures, working with recursive terms, reordering of relational program execution
and making relational queries over a non-constant number of variables in continuation-passing style. To implement some of them,
we suggested some helpful non-relational helpers for \textsc{miniKanren} implementation libraries: distinguishing between variable
and non-variable terms and ``occurs hooks''. We shown that the non-relational nature of suggested helpers may cause 
search incompleteness, but provided a way to address that drawback.

Despite the used optimizations we sometimes encounter problems with the number of branches in search tree that causes high time and
memory consumption. In the future we plan to address these problem and provide a more elaborated approach for a production-ready
solver for the given constraint system as well as other similar ones.

\bibliographystyle{plain}
\bibliography{main}

\end{document}